\documentclass[aps,pre,reprint, amsmath, amssymb,superscriptaddress]{revtex4-1}
\usepackage{graphicx} 
\usepackage{braket}

\pdfoutput=1
\usepackage[utf8]{inputenc}
\usepackage{bm}
\usepackage[english]{babel}
\usepackage[T1]{fontenc}
\usepackage{mathrsfs}
\usepackage[retainorgcmds]{IEEEtrantools}
\usepackage{amsmath}
\usepackage{amssymb}
\usepackage{color}
\usepackage{amsfonts}
\usepackage{times,txfonts}
\usepackage{nicefrac}
\usepackage[colorlinks=true,linkcolor=blue,urlcolor=blue,citecolor=blue,pdfusetitle]{hyperref}
\usepackage{mathtools}
\usepackage{isotope}
\usepackage{siunitx}
\usepackage{dsfont}
\usepackage[percent]{overpic}

\usepackage{float}
\graphicspath{{images/}}

\usepackage{braket}

\usepackage{lipsum}
\usepackage{enumerate}

\usepackage[dvipsnames]{xcolor}
\usepackage{accents}
\usepackage{diagbox}

\begin{document}
\setlength{\abovedisplayskip}{4pt}
\setlength{\belowdisplayskip}{4pt}

\title{Quantum computing architecture with trapped ion crystals and fast Rydberg gates}

\author{Han Bao}%
\email{hanbao@uni-mainz.de}
\affiliation{QUANTUM, Institut f\"{u}r Physik, Universit\"{a}t Mainz, D-55128 Mainz, Germany}%

\author{Jonas Vogel}%
\affiliation{QUANTUM, Institut f\"{u}r Physik, Universit\"{a}t Mainz, D-55128 Mainz, Germany}%


\author{Ulrich Poschinger}%
\affiliation{QUANTUM, Institut f\"{u}r Physik, Universit\"{a}t Mainz, D-55128 Mainz, Germany}%

\author{Ferdinand Schmidt-Kaler}%
\email{fsk@uni-mainz.de}
\affiliation{QUANTUM, Institut f\"{u}r Physik, Universit\"{a}t Mainz, D-55128 Mainz, Germany}%
\affiliation{Helmholtz-Institut Mainz, D-55128 Mainz, Germany}


\begin{abstract}
Fast entangling gate operations are a fundamental prerequisite for quantum simulation and computation. We propose an  entangling scheme for arbitrary pairs of ions in a linear crystal, harnessing the high electric polarizability of highly excited Rydberg states. An all-to-all quantum gate connectivity is based on an initialization of a pair of ions to a superposition of ground- and Rydberg-states by laser excitation, followed by the entangling gate operation which relies on a state-dependent frequency shift of collective vibrational modes of the crystal. This gate operation requires applying an electric waveform to trap electrodes. Employing transverse collective modes of oscillation, we reveal order of $\mu s$ operation times within any of the qubit pairs in a small crystal. In our calculation, we are taking into account realistic experimental  conditions and feasible electric field ramps. The proposed gate operation is ready to be combined with a scalable processor architecture to reconfigure the qubit register, either by shuttling ions or by dynamically controlling optical tweezer potentials. 
\end{abstract}

\maketitle


\section{\label{sec:1}Introduction}
Trapped atomic ions are among of the promising systems for building universal quantum computers and quantum simulators. The most widely used two-qubit entangling gates rely on state-dependent geometric phases, including M\o lmer-S\o rensen gate~\cite{MSgate} and light shift~\cite{Lightshiftgate} gates. The underlying mechanism relies on an oscillatory state-dependent force, leading to transient state-dependent displacement of the driven oscillation mode. A dominant limitation of the gate speed stems from the fact that parasitic excitation of other collective modes has to be suppressed. Therefore, typical gate times increase with the length of the ion crystal, exceeding 100~$\mu$s under typical operation conditions ~\cite{Monroe2014PRLgate}. On the other hand, quantum processors based on neutral atom registers employ Rydberg interactions and realize gate operations on orders of magnitude faster timescales~\cite{Lukin2023Nature}.  

To improve the speed of two-qubit operations in trapped ion processors, non-adiabatic schemes have been proposed~\cite{garciaripol,Steane_2014,PRA2020Shapira,Torrontegui_2020,shapira2023fast} and demonstrated~\cite{Monroe2014gate}. To date, the fastest gates achieved in a two-ion crystal are ranging from 1.6~$\mu s$ (99.8\% fidelity) down to 480~ns (60\% fidelity)~\cite{Lucasfastgate}. The design of laser pulses ensures that the trajectories of all modes are closed simultaneously at the end of the sequence. Now, limitations of the achievable gate speed are determined by the available laser power, and additional sophistication arises from the fact the the gate-mediating force becomes non-linear for fast gates~\cite{saner2023,Uli2010cat}. \\

In this work, we explore the options for harnessing the dipole-dipole interaction mediated by excitation to Rydberg states for new pathways in trapped ion quantum processors. An entangling gate within 700ns (78\% fidelity) has been demonstrated~\cite{RydgateZhang} on a two-ion crystal. However, due to  Coulomb repulsion, the inter-particle distance for trapped ions of a few $\mu$m are larger as compared to neutral atoms in optical lattices, reducing the dipole-dipole interaction, suppressing long-range interactions within a long ion crystals. We discuss how the range limitation can be circumvented with a Rydberg-enabled gate, allowing for all-to-all connectivity on a small crystal of ions. The scheme is based on state-independent electric forces to realize high fidelity, fast and all-to-all quantum gate operations, taking benefit from the high polarizability~\cite{Jonasgate,kick_measure_P} of the ions excited to Rydberg states. \\

This manuscript is organized as follows: First, we describe the all-to-all gate mechanism, taking a $^{40}Ca^+$ ion crystal and typical operating parameters as an example. Based on this, we describe a theoretical framework and present a detailed scheme for optimizing the electric waveforms for gate operations. We start with a scheme consiting of discrete consecutive electric kicks, and extend this to a continuous electric displacement scheme. Our scheme is based on purely electric forces acting in radial direction, perpendicular to the linear ion crystal. This has several advantages: The gate operation is driven with a weak electric field, far below the electric field strength, where Rydberg states in the Stark map would cross (Inglis-Teller limit). The maximal spatial excursion during the electric kick is sufficiently small to avoid any effect of trap field anharmonicity. Furthermore, the radial gate operation implemented between a pair of arbitrary two ions in a linear crystal of a few and can conveniently combined with shuttling operations, moving trapped-ion qubits into and out of the Rydberg excitation zone~\cite{Vidyut2017PRA,Hilder2022PRX,racetrack}. Alternatively, dynamical control of optical tweezer potentials \cite{Ozeri2024tweezer} could be employed.  

\begin{figure*}\label{Fig:schematic}
    \centering
    \includegraphics[width=0.95\linewidth]{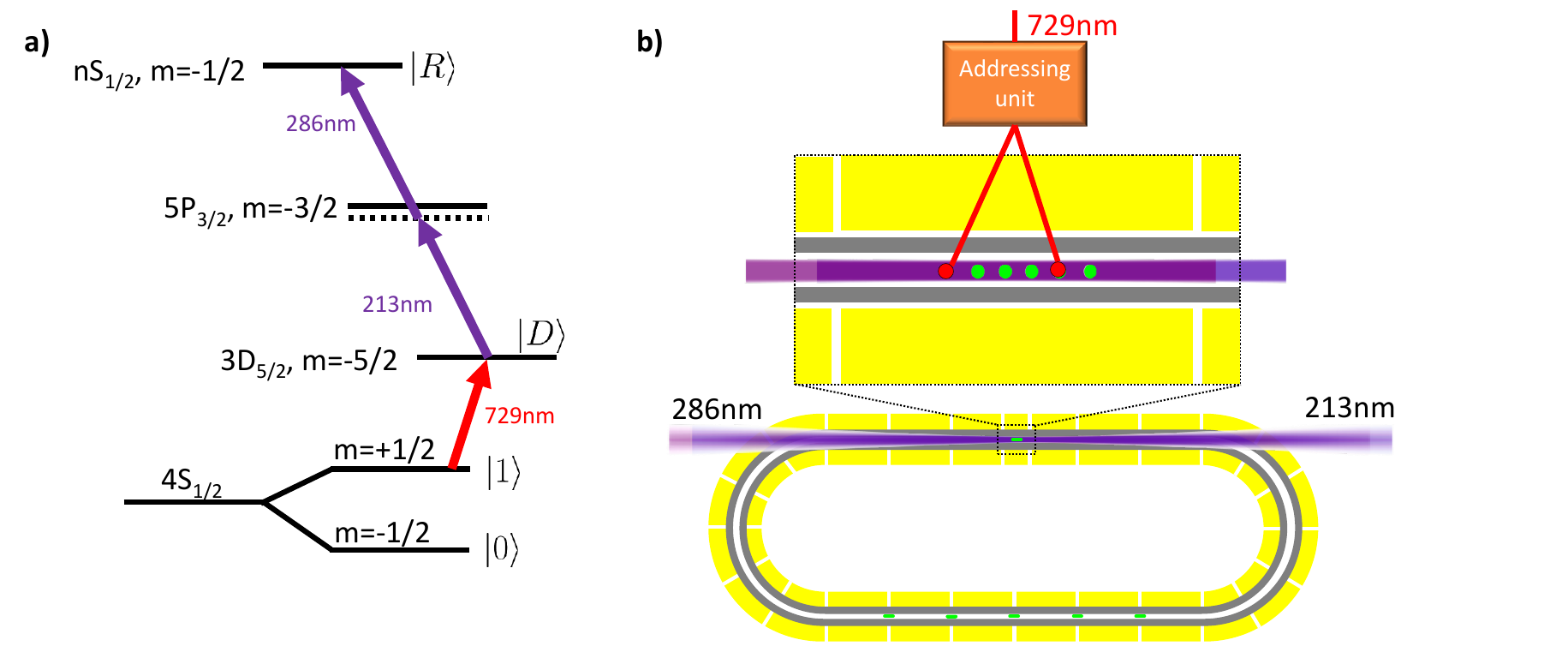}
    \caption{\textbf{(a)}: Relevant energy levels and transitions for $^{40}$Ca$^+$ ion. The qubit is encoded electronic levels, see text.
    \textbf{(b)}: Proposed architecture - crystals of trapped ions (green dots) are stored in different regions of the segmented racetrack Paul trap (rf electrodes in grey, dc electrodes in yellow) and can be shuttled between these. The shown six-ion crystal allows for addressed manipulation with 729~nm for selecting the qubits to undergo a two-out-of-N gate operations.  These ions are excited into the Rydberg state by counter-propagating focused beams near 286~nm and 213~nm, while all other ions in the interaction zone (green dots) are not excited to Rydberg and participate in the radial excursion of the linear crystal, but not in the quantum gate. More ions may be stored outside the laser interaction zone to realize a scalable architecture.}
\end{figure*}

The relevant energy levels for $^{40}$Ca$^+$ ions are shown in Fig.~\ref{Fig:schematic}(a). The qubit is encoded in the Zeeman sub-levels of the ground state  $\{|0\rangle,|1\rangle\}=|4S_{1/2}, m=\pm 1/2\rangle$. Population from the state $|1\rangle$ can be coherently transferred into the meta-stable state $|D\rangle=|3D_{5/2},m=-5/2\rangle$ with laser radiation driving a dipole-forbidden transition near 729nm. For a two-qubit gate on any pair of ions within a crystal, this transfer operation is performed by  addressing the ions to be gated with tightly focused laser beams near 729nm. Then, the population in the metastable  state is further excited into the Rydberg state $|R\rangle=|nS_{1/2}, m=-1/2\rangle$ using laser radiation near 213~nm and 286~nm. For each of the two gate ions, the excitation sequence coherently maps $\ket{1}$ to $\ket{R}$, as the van der Waals interaction strength of two ions at a typical distance of about 5$\mu$m is sufficiently weak for the Rydberg states which we discuss here. The trapped ions do not undergo a Rydberg blockade effect, which neutral atoms feature due to their strong Rydberg spin-spin interaction. After this mapping step, the entire ion crystal's radial motion driven by a sequence of electric kicks. The high polarizability of Rydberg states -  negligible for the low-lying states - gives rise to state-dependent secular frequencies and therefore to state-dependent radial motion. Now, population in the Rydberg state $|R\rangle$ is transferred back to $|1\rangle$ by the reverse excitation sequence. All spectator ions in the crystal can participate in the transient gate motion, but their internal state does not influence the acquired gate phases. Finally, for the readout of the qubit register, we excite selectively transfer population from $|1\rangle$ to the meta-stable state $|D\rangle$. Now, under resonant excitation near 397nm  and 866nm (not shown in Fig.~\ref{Fig:schematic}), the ions only emit resonance fluorescence when measured in $|1\rangle$ and not fluorescence otherwise. 

Note that the proposed scheme does not require individual ion addressing at UV wavelengths, but rather relies of addressing of ions for IR wavelength only~\cite{NAEGERL_PRA1999}, a proven technology. Also, the Rydberg excitation is carried out using focused, counter-propagating laser beams, directed along the axis defined by the linear ion crystal and with partial Doppler cancellation. Moreover, any recoil upon excitation to Rydberg states will be in axial, but not in the radial direction, and therefore not affect the gate operation. The ion crystal is aligned along the trap axis ($z$), while the electric gate drive field $E_x(t)$ is applied along one of the radial directions ($x$). Thus, in our proposed architecture, any parasitic axial excitation of the crystal, e.g. from ion shuttling operations in the segmented trap, does not affect the radial excitation, which is relevant for the gate operation. 

To describe the all-to-all gate operations in detail, we outline the theory of Rydberg-enabled state-dependent forces for ion crystals in the following.

\section{Theoretical framework}
\subsection{State-dependent secular trap frequencies}

For an ion in a radio frequency (rf) trap, the electric potential can be written as
\begin{equation}
\begin{aligned}
\Phi(\mathbf{r}, t)=&\gamma_{\mathrm{rf}} \cos \left(\Omega_{\mathrm{rf}} t\right)\left[x^2-y^2\right]\\
-&\gamma_{\mathrm{dc}}\left[(1+\epsilon) x^2+(1-\epsilon) y^2-2 z^2\right]
\end{aligned}
\end{equation}
where $\gamma_{dc}$ is the static field gradient, $\gamma_{rf}$ is the gradient of the rf field oscillating at drive frequency $\Omega_{rf}$, and the dimensionless parameter $\epsilon$ describes the radial asymmetry of the dc confinement. The secular oscillation frequencies of an ion of mass $M$ and charge $e$ are given by
\begin{equation}
\begin{aligned}
\omega_{x,y}^2 & =\frac{2 e^2 \gamma_{\mathrm{rf}}^2}{M^2 \Omega_{\mathrm{rf}}^2}-\frac{2 e}{M} \gamma_{\mathrm{dc}}(1\pm\epsilon), \\
\omega_{z}^2 & =\frac{4 e \gamma_{\mathrm{dc}}}{M} .
\end{aligned}
\end{equation}
If the ion is excited into a Rydberg state with nonzero polarizability $\alpha$, the secular frequencies are changed to be~\cite{Higgins}
\begin{equation}
\begin{aligned}
\tilde{\omega}_{x,y}^2 & =\omega_{x,y}^2-\frac{ \alpha}{M}\left[\gamma_{\mathrm{rf}}^2+2\gamma_{\mathrm{dc}}^2(1\pm\epsilon)^2\right], \\
\tilde{\omega}_z^2 & =\omega_{z}^2-\frac{16 \alpha}{M} \gamma_{\mathrm{dc}}^2.
\end{aligned}
\end{equation}
As we show in the following, such state-dependent secular trap frequencies can lead to coupling between the ion's electronic state to their collective motion, which can be leveraged to generate entangling gates.

As typical trap operation parameters we assume $\{\gamma_{\text{DC}},~\gamma_{\text{RF}},~\epsilon,~\Omega_{\text{RF}}\}=\{6.406\times 10^7V/m^2,~1.123\times 10^9V/m^2,~0.400,~2\pi\times 14.135\text{MHz}\}$, leading to bare secular trap frequencies $\{\omega_{x},\omega_{y},\omega_{z}\}=2\pi\times\{6.000,~6.500,~3.953\}~\text{MHz}$. With the static  polarizability $\alpha\approx 1.3\times 10^{-30}C\cdot m^2V^{-1}$ for the $^{40}$Ca$^+$ Rydberg energy level $49S_{1/2}$~\cite{kick_measure_P}, the frequency change for a single confined ion in transverse $x$-direction is $2\pi\times 53~\text{kHz}$, but merely $2\pi\times 2.4~\text{kHz}$ for the $z$-direction.

\subsection{Linear ion crystals with state-dependent modes of oscillation}
For a linear chain of $N$ ions, the properties of the motional modes can be derived by expanding the state-dependent potential energy, including electrostatic and pondermotive trap potentials and Coulomb repulsion, up to second order around the ion equilibrium positions. For each direction $j=x,y,z$, the motion is described by a Hessian matrix $B^{(j)}$~\cite{Marquet2003}, which depends on the internal state of each ion $\ket{s}$,with $s=0$ for the ground state $\ket{0}$ and $s=1$ for the highly excited Rydberg state $\ket{R}$. In a common radial excitation, all ions in the crystal participate but only those, which have been excited in the Rydberg state, do show a state dependent trajectory. We indicate them by a specific $m$ with s$_m$=1. The state-dependent \emph{local} secular frequency of ion $m=1\hdots N$ reads as 
\begin{equation}
\begin{aligned}
\omega_{jm}=\begin{cases}\omega_{j}&\text{if $s_m=0$}\\
\tilde{\omega}_j&\text{if $s_m=1$}
\end{cases}
\end{aligned}
\end{equation}
The $B^{(j)}$ matrices are 
\begin{equation}
\begin{aligned}
B^{(z)}_{mn}=&\begin{cases}\frac{\omega_{zm}^2}{\omega_{z}^2}+2 \sum_{\substack{p=1 \\ p \neq m}}^N \frac{1}{\left|u_m-u_p\right|^3} & \text { if } m=n, \\ \frac{-2}{\left|u_m-u_n\right|^3} & \text { if } m \neq n,\end{cases}\\
\\
B^{(x,y)}_{mn}=&\frac{2\omega_{(x,y)m}^2+\omega_{zm}^2}{2\omega_{z}^2}\delta_{m n}-\frac{1}{2} B^{(z)}_{mn}
\end{aligned}
\end{equation}
with $u_n$ being the dimensionless equilibrium position of ion $n$ defined by $u_n=z_n^{(0)}/l$,  where $z_n^{(0)}$ is the equilibrium position of  ion $n$ and the length scale $l^3=e/(4\pi\epsilon_0M \omega_z^2)$. Note that the bare secular frequency $\omega_{z}$ is used to calculate the equilibrium positions. Again, the very weak van der Waals interaction strength between two ions at a typical distance of 5~$\mu$m allows for this approximation. The properties of the secular modes follow from the eigenvalues and eigenvectors of the $B^{(j)}$ matrices:
\begin{equation}
\begin{aligned}
B^{(j)}\vec{b}_{k}^{(j)}=\mu_{k}^{(j)}\vec{b}_{k}^{(j)} \qquad k=1\hdots N
\end{aligned}
\end{equation}
The components $b_{kn}^{(j)}$ of the normalized eigenvectors describe the relative oscillation amplitude of ion $n$ upon excitation of mode $k$ along direction $j$. The mode frequencies $\nu_{k}^{(j)}$ are given by corresponding eigenvalues $\mu_{k}^{(j)}$ via
\begin{equation}
\begin{aligned}
\nu_{k}^{(j)}=\sqrt{\mu_{k}^{(j)}}\omega_{z}
\end{aligned}
\end{equation}
Note that both the secular frequencies and the mode eigenvectors depend on the total internal state of the ion crystal. From now on, we consider only the transverse $x$-direction and drop the superscripts denoting the direction. State-dependent mode frequencies for a two-ion crystal are shown in Table~\ref{table:1}.
\begin{table}[h!]
   \begin{center}
        \par\medskip
        \begin{tabular}{|c|c|c|c|} 
        \hline
        \diagbox{Motional}{$|s_1s_2\rangle$}& $|0~0\rangle$ & $|0~R\rangle$, $|R~0\rangle$ & $|R~R\rangle$\\
        \hline
        COM mode freq $\nu_{1}^{(s_1s_2)}/2\pi$(MHz) & 6.000 & 5.974  & 5.947\\
        \hline
        Rock mode freq $\nu_{2}^{(s_1s_2)}/2\pi$(MHz) & 4.514 & 4.478  & 4.443\\
        \hline
        \end{tabular}
        \caption{Example values of state-dependent transverse oscillation frequencies $\nu_{k}^{(s_1s_2)}$ for a two-ion crystal, for all internal basis states $\ket{s_1s_2}$ and for the in-phase (COM) and out-of-phase (Rock) oscillation modes.}\label{table:1}
   \end{center}
\end{table}

\begin{figure*}\label{Fig:4-kick}
    \centering
    \includegraphics[width=0.95\linewidth]{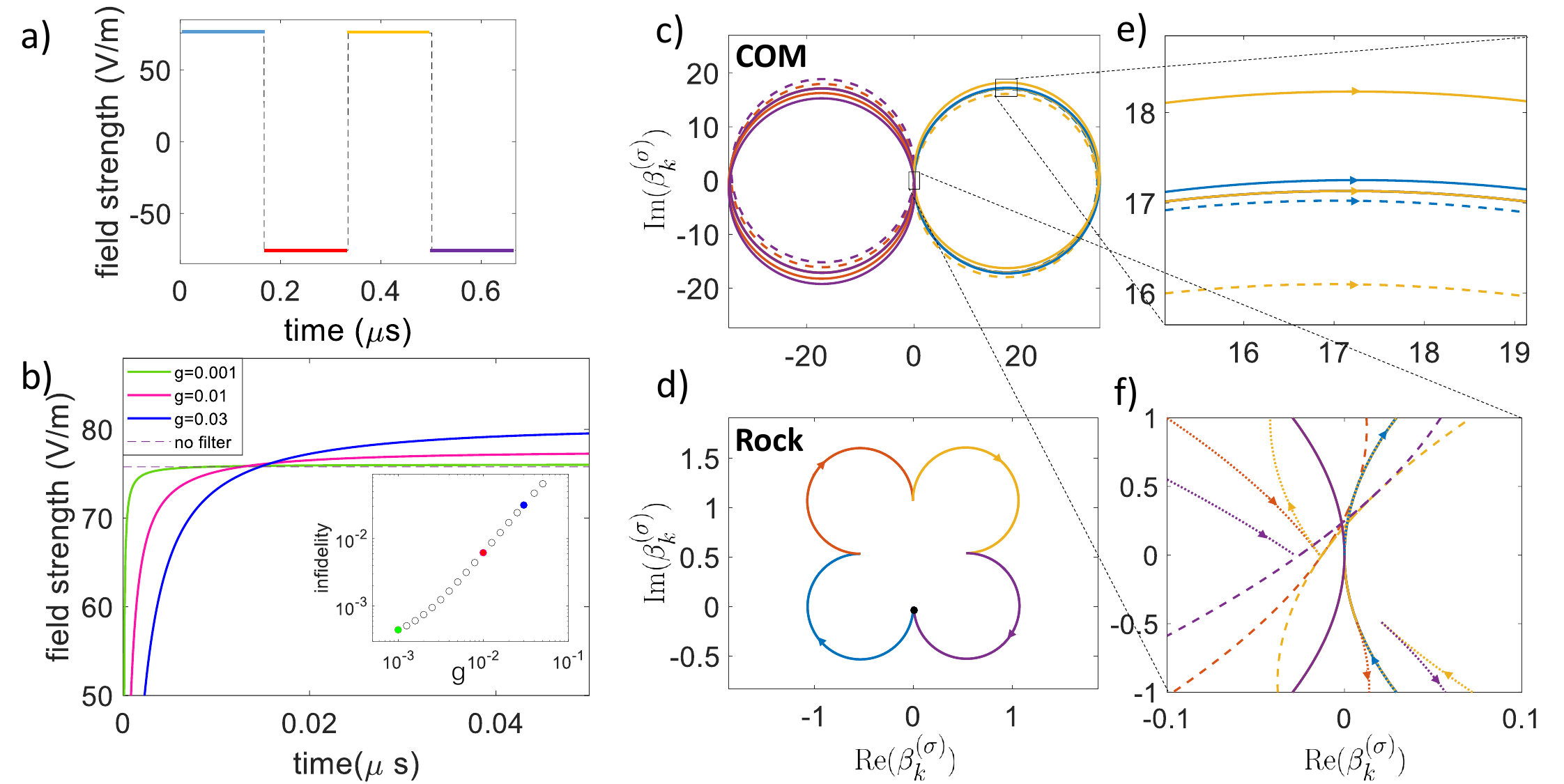}
    \caption{\textbf{Discrete four-kick scheme for two ions}. The trap parameters chosen as in Tab.~\ref{table:1} and to fulfill frequency multiple relation $4\nu_2^{(0R/R0)}=3\nu_1^{(0R/R0)}$. \textbf{(a)}: Kicking waveform, such that the gate time is $t_g=(2/\pi){\nu_{1}^{0R}}\times4=0.67\mu s$. The parts of each trajectory in (c)–(f) correspond to the four kicks (colors scheme). \textbf{(b)} Square pulse distortion induced infidelity: The dashed line is the square wave without distortion, while the three solid lines are function $G(t)=(2A/\pi)\arctan(\sin(2\nu_{1}^{0R}t)/g)$ with different values of $g$. The amplitude $A$ for each curve is compensated to fulfill $\Delta \phi=\pi$. The inset shows the resulting infidelity induced by the distorted square wave pulse. The three solid circles with corresponding the colors represent the three values of $g$. \textbf{(c)} The COM mode motional trajectories, for the internal states $|0R\rangle$ and $|R0\rangle$ (solid lines). COM mode trajectories are closed since the gate time $t_g$ is the integer multiple of oscillation period $(2/\pi){\nu_{1}^{0R}}$. For the internal states $|00\rangle$ (dashed line) and $|RR\rangle$ (dotted line), the COM mode trajectories are not closed. The residual displacement and phonon number is small, $3.7\times 10^{-4}$, and induces 0.04\% infidelity. \textbf{(d)} The rock mode motional trajectories: for the internal states $|00\rangle $ and $|RR\rangle$, the rock mode is not excited due to the symmetry of the two ions (black dot at the zero point). For the internal states $|0R\rangle$ and $|R0\rangle$ (solid lines), the rock mode is excited but the trajectories are closed due to frequency multiple relation. \textbf{(e)} and \textbf{(f)} are zoom-in view of \textbf{(c)}.
    }
\end{figure*}

\section{Transverse electric kick-induced entangling gates}
Now we consider applying a spatially homogeneous, time-dependent electric force $f(t)$ along the $x$-direction. We limit the treatment to a two-ion crystal with internal states $\ket{\sigma}=\{\ket{0,0},\ket{0,R},\ket{R,0},\ket{R,R}\}$ with oscillation frequencies $\nu_k^{(\sigma)}$. The time evolution of the motional state of the crystal for a fixed internal state $\ket{\sigma}$ is described by the propagator
\begin{equation}
U^{(\sigma)}_{\mathrm{I}}(t)=\prod_{k=1}^N\mathcal{D}_k\left[\beta_k^{(\sigma)}(t)\right] \exp \left[\mathrm{i} \varphi_k^{(\sigma)}(t)\right]
\end{equation}
with the displacement operator $\mathcal{D}_k$ of mode $k$ and state-dependent displacements
\begin{eqnarray}
\beta_k^{(\sigma)}(t) & =&-\frac{i}{\hbar}\int_{0}^t F_k^{(\sigma)}(\tau) e^{-\mathrm{i} \nu_{k}^{(\sigma)} \tau} \mathrm{d} \tau 
\end{eqnarray}
The acquired state-dependent geometric phase from mode $k$ reads
\begin{eqnarray}
\varphi_k^{(\sigma)}(t) & =&\frac{1}{\hbar^2}\int_{0}^t \int_{0}^{\tau} F_k^{(\sigma)}\left(\tau^{\prime}\right) F_k^{(\sigma)}(\tau) \nonumber \\
& \times &\sin \left[\nu_{k}^{(\sigma)}\left(\tau^{\prime}-\tau\right)\right] \mathrm{d} \tau \mathrm{d} \tau^{\prime}
\end{eqnarray}
Here, the net force on mode $k$ is $F_k^{(\sigma)}(t)=f(t)W_{k}^{(\sigma)}l_{k}^{(\sigma)}$ with  $l_{k}^{(\sigma)}=(\hbar/M\nu_{k}^{(\sigma)})^{1/2}$ being the characteristic length and  $W_{k}^{(\sigma)}=\sum_{n}b^{(\sigma)}_{kn}$. All motional modes have to to be restored to rest at the end of the gate action $t_g$, in order to avoid residual entanglement between internal states and motion. A maximally entangling controlled-Z gate
\begin{equation}
CZ= \text{diag}(1,1,1,-1)
\end{equation}
requires a $\pi$ phase shift between $|R,R\rangle$ and the other basis states, up to local equivalence. These requirements read
\begin{align}
\label{condition1}&\beta_k^{(\sigma)}(t_g)=0,  ~~\forall~k,~\sigma\\
\label{condition2}&\Delta\varphi=\varphi^{11}(t_g)+\varphi^{00}(t_g)-\varphi^{01}(t_g)-\varphi^{10}(t_g)=\pi
\end{align}
with the total geometric phase $\varphi^{(\sigma)}(t)=\sum_{k=1}^N \varphi_k^{(\sigma)}(t)$.
In the following, we describe a simplified scheme to provide a pictorial understanding of the gate dynamics for two ions. This few-parameter scheme allows for a straightforward optimization method. Crucially, the axial trap frequency is tuned to render the in-phase (COM) and out-of-phase (Rock) mode frequencies to be commensurate. Here, we choose $4\nu_2^{(0R,R0)}=3\nu_1^{(0R,R0)}$, see Tab.~1. Four kick pulses of duration $\tau=(2/\pi){\nu_{1}^{(0R)}}$ and constant field strength but alternating field direction are applied. In this way, the evolution of both modes in phase space  leads to closed trajectories for internal states $|0R\rangle$ and $|R0\rangle$. The Rock mode is not excited for $|00\rangle$ and $|RR\rangle$ as the forces on both ions cancel out. The trajectory of the COM mode for $|00\rangle$ and $|RR\rangle$ is not completely closed because of a slight frequency difference between $\nu_{1}^{(00)}$, $\nu_{1}^{(RR)}$ and $\nu_{1}^{(0R)}$. Correspondingly, a residual phonon number $5\times10^{-4}$ is calculated, see Fig.~\ref{Fig:4-kick} (f). Under typical experimental conditions $\nu_{1}^{(0R)}=2\pi\times5.974\text{MHz}$ and a gate time of 0.67$\mu s$. We observe, that the motional excitation is predominantly in the COM mode, as compared to the Rock mode.  \\
The feasibility of such discrete kick scheme is limited by the unrealistic assumption of infinitely steep flanks of the electric field pulses. In practice, the finite switching times of the electric field waveforms compromise the control accuracy required to suppress undesired residual state-dependent motion. To evaluate such effect, we use the function $G(t)=\frac{2A}{\pi}\arctan(\sin(2\nu_{1}^{0R}t)/g)$ to model electric field pulses with finite switching time. The parameter $g$ indicates how sharply the waveform rises and falls. The edges are sharper when $g$ is smaller. The variable amplitude $A$ is re-scaled to provide a constant pulse area for varying $g$. We find a reduction of gate fidelity for non-ideal electric pulses, see Fig.~2(b). \\
To avoid problems with switching behavior and allow for the state-dependent kick scheme to operator on larger ion chains, \emph{continuous} electric field waveforms can be employed. In the following, we present two different methods to compute control solutions, fulfilling the conditions for gate operation Eqs.~\eqref{condition1} and \eqref{condition2}.

\begin{figure*}
    \centering
    \includegraphics[width=0.95\linewidth]{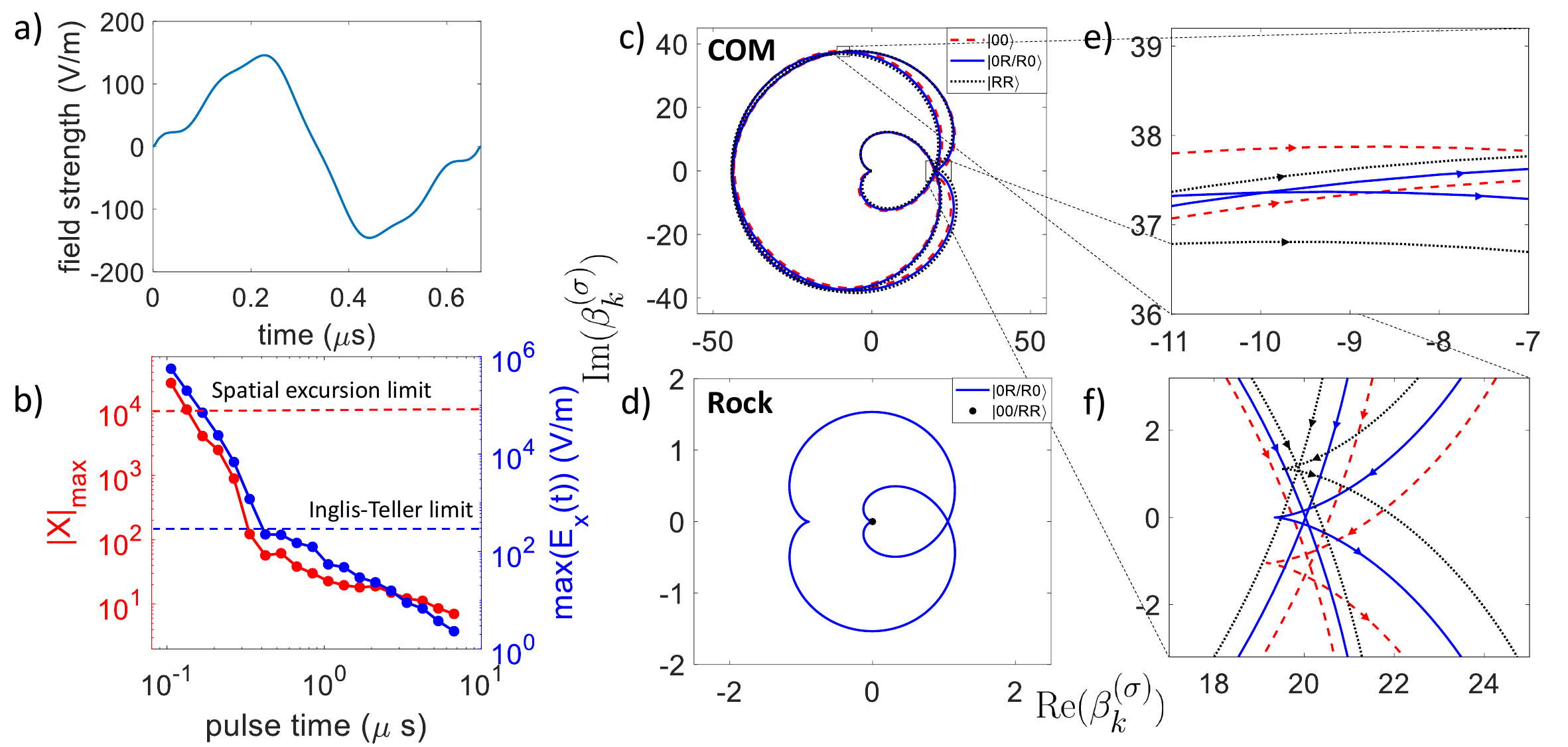}
    \caption{\textbf{Continuous-kick scheme for two ions} \textbf{(a)} Optimized continuous waveform for two ions. The trap frequencies and gate time $t_g$ are identical to that in Fig.~2. \textbf{(b)} Maximal field strength $|E(t)|_{max}$ and maximal spatial excursion $|X|_{max}$ as varying $t_g$. The dashed lines mark the Inglis-Teller limit of less than 10$\%$ admixture (blue) of 49S to the corresponding P state, and the estimated spatial excursion limit (red) of the trap. \textbf{(c)} Phase space trajectories of COM mode for the internal states $|00\rangle$ (red dashed line), $|RR\rangle$ (black dotted line), $|0R\rangle$ and $|R0\rangle$ (both blue solid lines). \textbf{(d)} Phase space trajectories of rock mode for all internal states. For the internal states $|00\rangle $ and $|RR\rangle$, the rock mode is not excited due to the symmetry of the two ions (black dot at the zero point). \textbf{(e)} and \textbf{(f)} are zoom-in view of \textbf(c). 
    }\label{fig:continuous-kick}
\end{figure*}

\section{Continuous-kick schemes}
\subsection{Time-domain control}\label{multi-kick}
We now divide the electric kick into $n_t$ slices of equal duration and amplitudes $f_n$. Slice $n$ starts at $t_{n-1}$ and ends at $t_n=nt_g/n_t$. The complete electric kick starts at $t_0=0$ and ends at $t_{n_t}=t_g$. Then, the closure condition Eq.~\eqref{condition1} can be written as
\begin{equation}
\begin{aligned}\label{condition expanded}
\beta_k^{(\sigma)}(t_g)&=\sum_{n=1}^{n_t}f_n\beta_{k,n}^{(\sigma)}=0,  ~~\forall~k,~\sigma
\end{aligned}
\end{equation}
with
\begin{equation}
\begin{aligned}
\beta_{k,n}^{(\sigma)}=W_k^{(\sigma)}l_{k}^{(\sigma)}\int_{t_{n-1}}^{t_n} e^{-\mathrm{i} \nu_{k}^{(\sigma)} \tau} \mathrm{d} \tau
\end{aligned}
\end{equation}
Eq.~\eqref{condition expanded} can be expressed as
\begin{equation}\label{matrixform}
\begin{aligned}
 \begin{bmatrix}
\text{Re}\;\beta_{k,n}^{(\sigma)}\\ \text{Im}\;\beta_{k,n}^{(\sigma)}
\end{bmatrix}\cdot\vec{f}=0\\
\end{aligned}
\end{equation}
The size of the matrix in Eq. \eqref{matrixform} is $8N\times n_t$. $N$ denotes the number of ions and also the number of modes in the $x$-direction. The factor 8 results from four possible configurations for $\sigma$ and the conditions for both real and imaginary parts of $\beta_{k,n}^{(\sigma)}$. The waveform samples are arranged as a vector $\vec{f}=\{f_1,...,f_n\}$. All solutions of Eq. \eqref{matrixform} fulfill the closure conditions. In the next step, a final control solution $\vec{f}_{opt}$ is built from a linear combination of the solutions Eq. \eqref{matrixform}, which ensures the accumulation of the required gate phase as per condition Eq. \eqref{condition2} at the minimum possible kick amplitude. To enforce that the electric field waveform to start and end at 0 V/m, we can make use of the symmetric or anti-symmetric properties of the most effective waveforms. A detailed description is presented in Appendix A. For a sufficient large number of slices $n_t$, the vector $\vec{f}_{opt}$ converges a quasi-continuous waveform. 

\subsection{Frequency-domain control}
In the following, we will be casting the waveform in spectral components~\cite{Ozeri2023PRL} to reduce the computation effort and the parameter space - thus converging quickly to the optimum. Moreover, a solution expanded in its Fourier components automatically becomes continuous. This ensures the experimental feasibility. We decompose the electric kick waveform into $n_f$ Fourier components, with frequencies $\omega_n$. The linear equations are
\begin{equation}
\begin{aligned}\label{Fourier condition expanded}
\beta_k^{(\sigma)}&=\sum_{n=1}^{n_f}\left[A_n^{(s)}W^{(\sigma)}_{k}l_{k}^{(\sigma)}\int_{0}^{t_g} \sin(\omega_n\tau)e^{-\mathrm{i} \nu_{k}^{(\sigma)} \tau} \mathrm{d} \tau\right]\\
&+\sum_{n=1}^{n_f}\left[A_n^{(c)}W^{(\sigma)}_{k}l_{k}^{(\sigma)}\int_{0}^{t_g} \cos(\omega_n\tau)e^{-\mathrm{i} \nu_{k}^{(\sigma)} \tau} \mathrm{d} \tau\right]\\
&=0,  ~~\forall~n,~\sigma
\end{aligned}
\end{equation}
$A_n^{(s/c)}$ are the amplitudes for sine and cosine waves. To ensure the sine waves starting and ending at zero V/m, we choose the frequency to be $\omega_n=n\pi/t_g$ with $n=1,2,...,n_f$. To ensure  all the cosine waves starting and ending at zero V/m, we give the equations an additional constrain as
\begin{equation}\label{cosine constrain}
\begin{aligned}
\sum_{n=1}^{n_f}A_n^{(c)}=0
\end{aligned}
\end{equation}
Eq. \eqref{Fourier condition expanded} and Eq. \eqref{cosine constrain} can be expressed as 
\begin{equation}\label{sinmatrixform}
\begin{aligned}
 \begin{bmatrix}
Re(\beta_{k,n}^{(\sigma)(s)})&Re(\beta_{k,n}^{(\sigma)(c)})\\ Im(\beta_{k,n}^{(\sigma)(s)})&Im(\beta_{k,n}^{(\sigma)(c)})
\end{bmatrix}\cdot\vec{A}=0\\
\end{aligned}
\end{equation}
where
\begin{equation}
\begin{aligned}
\beta_{k,n}^{(\sigma)(s)}&=W_k^{(\sigma)(s)}l_{k}^{(\sigma)}\int^{t_g}_{0} \sin(\omega_n\tau)e^{-\mathrm{i} \nu_{k}^{(\sigma)(s)} \tau} \mathrm{d} \tau\\
\beta_{k,n}^{(\sigma)(c)}&=W_k^{(\sigma)(s)}l_{k}^{(\sigma)}\int^{t_g}_{0} \cos(\omega_n\tau)e^{-\mathrm{i} \nu_{k}^{(\sigma)(s)} \tau} \mathrm{d} \tau\\
\vec{A}&=\begin{bmatrix}
\{A_{n}^{(s)}\}\\
\{A_{n}^{(c)}\}\\
\end{bmatrix}
\end{aligned}
\end{equation}
The size of the matrix is $8N\times 2n_f$. The length of the column vector $\vec{A}$ is $2n_f$. Then, the same method as in \ref{multi-kick} can be used to derive the best waveform. For sufficiently large $n_f$, the waveform converges to be the sames as derived in \ref{multi-kick}, see Appendix B.

\subsection{Continuous waveform results}
We start with gate operation on two ions only, see Fig.~\ref{fig:continuous-kick}(a)(c), in order to compare the continuous-kick scheme with the four-kick scheme, and choose the trap frequencies and gate time $t_g$ to be identical. Now, all trajectories are completely closed, ensuring that there is no fidelity loss by residual motional excitation. The waveform of the electric field is continuous and starts and ends at zero. This may mitigate problems with waveform distortions caused by an electronic filter functions, a limited amplifier slew rate or an inductance. Another advantage of the continuous-kick scheme is, that the gate time $t_g$ does not need to be integer multiple of oscillation period. For arbitrary $t_g$, condition Eq.~\eqref{condition1} can be  fulfilled by solving the linear equations \eqref{condition expanded} or \eqref{Fourier condition expanded}. Extra constrains can be added to the linear equations to ensure the waveform to start and end at 0, see appendix for details.

In the specific example with 0.67 $\mu s$ gate time, we find a maximum of required field of about 150~V/m, well below IT limit.  The peak value of the electric field $|E(t)|_{max}$ needed for different gate times $t_g$ is shown in Fig.~\ref{fig:continuous-kick}(c) (blue). Also, the ions' maximal spacial excursion in all trajectories of all possible motional modes and internal states is another important feature of our proposal, as shown in Fig.~\ref{fig:continuous-kick}(c) (red). We denote it as
\begin{equation}
|X|_{max}=max_{\sigma,k,t}(|Im(\beta^{(\sigma)}_k(t))|)
\end{equation}
The shorter the gate times $t_g$ we aim for, the higher becomes $|E(t)|_{max}$ and $|X|_{max}$, see Fig.~\ref{fig:continuous-kick} (c). We observe for this numerical simulation that there are two regimes: for longer pulse times (> 300ns) with a modest slope of $|E(t)|_{max}$, and correspondingly  phase space trajectories display at least two turns. However, in short pulse regime, the required field strength shows a much steeper rising slope. Here, the phase space trajectories consists of a sequence of many fractional circles, at the expense of high electric kick amplitudes. The onset of this regime is coinciding with a pulse time $t_g < 2 T_{trap}$, where $T_{trap}$ denotes the period of the radial confinement. As the proposed gate operates on the state-depending common modes of vibration, it is natural that this periodicity marks its speed limit. 

\section{Entangling arbitrary pairs of ions in a linear crystal}

\begin{figure}
    \centering
    \includegraphics[width=0.80\linewidth]{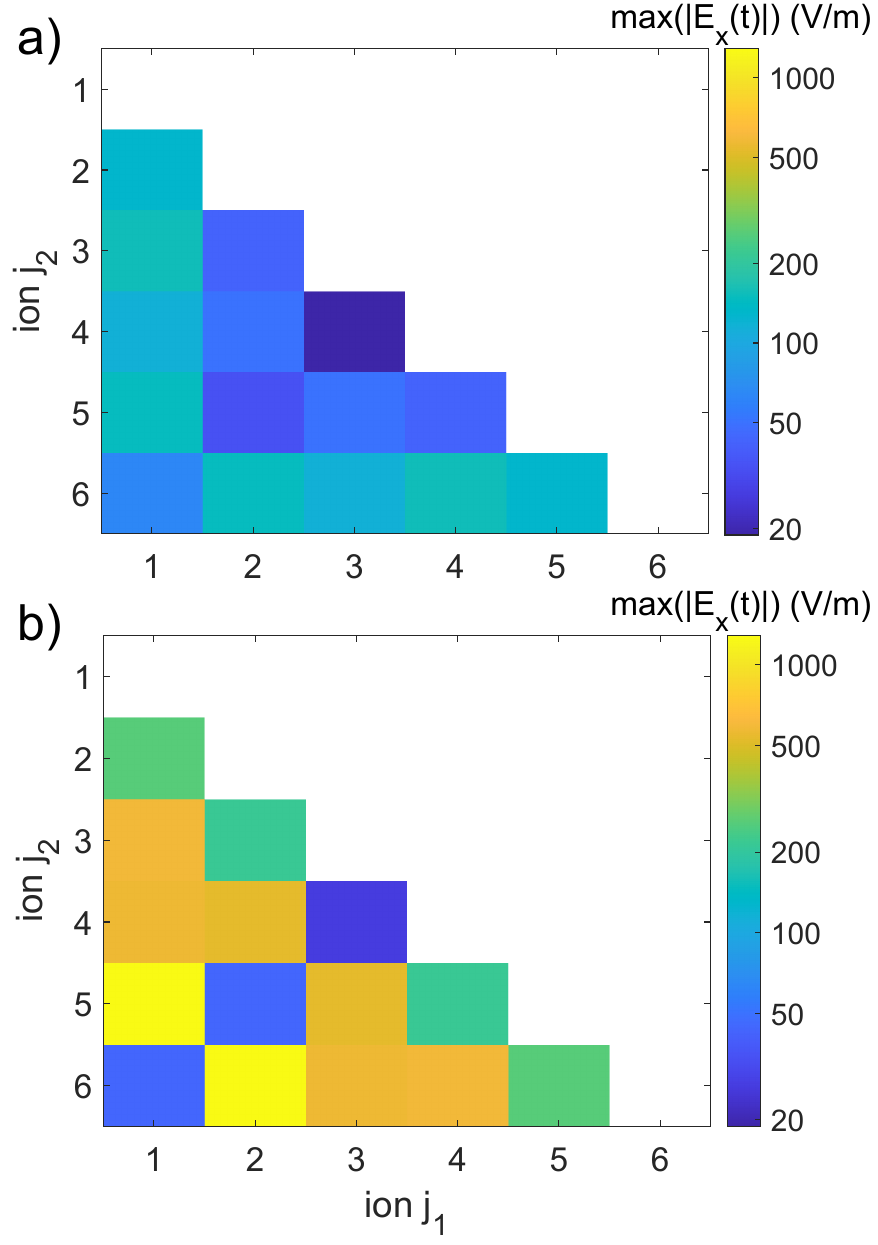}
    \caption{The maximal electric field to entangle arbitrary two ions in a six ion crystal. All 15 possible choices are calculated. The gate time is set to be \textbf{(a)} For n=49s, and gate time of 2.5 $\mu s$, the maximum electric field needed is below Inglis-Teller limit for all interactions. \textbf{(b)} But reducing the gate time to 2$\mu s$, in some of the cases e.g. where ion positions are far apart, the interaction would demand a much higher maximum electric field and would exceed the IT limit and appear not experimentally feasible. 
    }\label{fig:2in6}
\end{figure}

For the gate between an arbitrary pair of two ions out of a crystal with N ions, we take N=6 as concrete example, see Fig.~1(b). In such situation, six motional modes in the kicking direction all contribute to the entanglement. We denote the index of the two ions we want to entangle as $j_1$ and $j_2$. To evaluate the effective 1-to-1 interaction strength, the maximal electric field needed $|E(t)|_{max}$ is calculated for all 15 possible choices of ions and a gate time of 2.5 $\mu$s, see Fig.~\ref{fig:2in6}(a). The weaker the interaction is, the stronger the electric field required for the entangling gate, and the more motional displacement needs to be excited for the gate operation.

The effective 1-to-1 interaction strength depends on ion distance, ion crystal symmetry and trap anisotropy: Firstly, the interaction becomes weaker as ion distance increases. That's because for closer Rydberg ions, some of the the mode frequencies change more, as compared to more distant Rydberg ions. Thus, a state dependent phase is accumulated more effectively.  

Secondly, for a symmetric choice of ions ($j_1=N+1-j_2$), all motional modes except COM for $|RR\rangle$ state are not excited, reducing the constrains of condition Eq.~\eqref{condition1}. Thus, entangling symmetric ion crystal requires much weaker electric field, see the diagonal entries in Fig.~\ref{fig:2in6}. Finally, the trap anisotropy, expressed by a parameter $\kappa=\omega_z^2/\omega_x^2$, also affects the interaction: If $\kappa$ is increased, the axial constrain becomes more tightly than the radial constrains and the ion crystal will finally turn into zigzag structure. At the critical point of this transition, the trap anisotropy parameter is denoted as $\kappa_{crit}$. In such a critical point, the effective 1-to-1 interaction becomes the strongest. That's because the mode frequency $\nu^{(\sigma)}_{N}$ approaches  0 and creates the relatively largest mode frequency change.

\section{Assessment of the novel gate scheme and scaling properties}
The gate operation is based on state-dependent trajectories, similar to commonly used gate operations e.g. with optical forces in polarization gradient formed by two beams of different polarization. The interaction in such a traveling standing wave is proportional to $(a+a^{\dagger})$ only if the phonon number is sufficiently small, such that higher order terms can be neglected, this is coined the Lamb Dicke regime. Contrary, outside this regime, phase space orbits are no longer circles~\cite{Uli2010cat,Lucasfastgate} and correspondingly, a residual motional excitation may reduce the gate fidelity. Back to the kicking gate operation on Rydberg states, we state that the forces are linear in this case, even for very high motional excitation, limited by two effects: First, the electric field between the segments of the linear Paul trap is linear to a high degree, as long as the excursions in space are small as compared to the distance between the ions and the electrode surface. The spacial excursion $|X|_{max}$ measures the excursion in length scales of the motional ground state of the wave packet, about 10~nm here. Assuming the linear region for typical Paul trap is $100\mu m$, the spatial excursion limit is $|X|_{max} < 10^4$. The second and more stringent limit occurs, when the electric field strength approaches the value to mix the initial Rydberg state with other states in the Stark map (Inglis-Teller limit). In our example, we choose 49S state since its polarizability has been measured accurately~\cite{kick_measure_P}. For the $49S$ and a gate time of 0.67 $\mu s$, the maximum electric field of 150~V/m required accounts for 97\% purity and only 3\% admixtures of the next Rydberg P state. 

To entangle arbitrary two ions out of an ions crystal with six ions, the required  gate time increased as compared to the situation of a two ion crystal. This is, because more constrains need to be fulfill for a closure of trajectories. We show the result of optimization in Fig.~\ref{fig:2in6} (a), choosing a gate time of 2.5$\mu s$.  The electric field needed is well below the Inglis-Teller limit and we see that the scheme will be realistic in all cases of the all-to-all gate operation. But just a small reduction of the chosen gate time to 2$\mu s$ results in much higher electric fields. 

Fast gate operations are essential, because eventually, after optimizing the kick waveform to close all trajectories, the Rydberg state lifetime becomes the dominating factor for the gate fidelity. Taking into account the 49s lifetime of 6.2 $\mu s$, the fidelity of a 0.67$\mu s$ gate would be reduced to 81\% only because of the lifetime. If we reduce the gate time to 0.1 $\mu s$, the infidelity from this reason could be reduced to about 3\%. But such fast operation would require an increased radial trap frequency. The better option might be to excite ions into the 49P state which exhibits a much longer lifetime of 190 $\mu s$. Now, the infidelity would become 7\% for 0.67 $\mu s$ gate time and drop to 0.1\% for 0.1 $\mu s$. For the entangling of two ions out of an ion crystal in 2.5 $\mu s$ gate time using the 49P state we expect a fidelity loss of 3\% due to spontaneous decay. Note, that the Rydberg state lifetimes are reduced the black body radiation, thus one may require a cryogenic operated system.


Let us conclude the assessment with a discussion of the scaling properties: As the Rydberg polarizability increases with the principal quantum number n like n$^7$, but the Inglis-Teller limit scales only with n$^{-5}$,  using a higher n will always improve the gate dynamics: As a consequence, lower electric fields are required for a fixed gate time, when increasing the principle quantum number $n$. Alternatively, one might reduce the gate time, when choosing a higher $n$. However, the gate time can not be reduced beyond a certain limit. If we assume  an electric field strength to keep 90\% state purity (IT limit), the minimal gate time converges to 0.38 $\mu s$ for large n, taking our trap parameters into account. Only higher trap frequencies would change this limit. 
But even though the gate times may not be reduced, a higher $n$ is of advantage, since the longer lifetime of the state, scaling with n$^3$, is leading an improved gate fidelity.

\section{Outlook}
We have presented a gate operation scheme using Rydberg ions and shaped electric fields waveforms. Taking realistic parameters into account, we find operation times of few $\mu$-seconds in an all-to-all situation with qubits arranged in a linear crystal. In future, one may explore the scheme experimentally: First, one might implement kicks on Rydberg ions in a similar fashion like demonstrated  in~\cite{FSK2012PRL,Bowler2012PRL}. A next step would be to explore the stability of Rydberg states under such high electric field waveforms and to monitor the phase space trajectories of a single Rydberg ion first, taking advantage of tools which have been developed for observing the Schrödinger's cat states of internal spin and external motional states~\cite{Uli2010cat}. 

For trapped ion quantum processors, the main focus today is scalabilty and fidelity of gate operations. Eventually, when algorithms will address more advanced problems and the corresponding circuits become much deeper, the speed of gate operations will become more relevant and raise the interest in faster gates,  taking benefit of Rydberg physics in trapped ions. 

\begin{acknowledgments}
We thank Roee Ozeri and Nitzan Akerman for discussions, we acknowledge the German Research Foundation (DFG) within the frameworks of the Forschergruppe GiRyd and Sonderforschungsbereich TRR-306 for support.
\end{acknowledgments}

\section{appendix}
Here we present the linear algebra method to derive the best waveform for the multi-kick and multi Fourier components respectively.
\subsection{Deriving best waveform for multi-kick method}
To optimize $|\Delta\varphi|$ for arbitrary normalized vector $\vec{f}$ in the null space, we denote the basis of the null space as $\vec{b}_k$. Any vector in the null space can be expressed as
\begin{equation}
\begin{aligned}
\vec{f}=\sum^j c_j\vec{b}_j=b\cdot\vec{c}
\end{aligned}
\end{equation}
where $c_k$ is the combination coefficient and forms the vector $\vec{c}$. $b$ is the matrix with each volume being $\vec{b}_k$.

We define a square matrix $P^{(\sigma)}$ to be
\begin{equation}
\begin{aligned}
P_{k_1, k_2}^{(\sigma)}=&\sum^k\Bigg[\int_{t_0}^t \int_{t_0}^{\tau^{\prime}} b_{k_2}(\tau^{\prime}) b_{k_1}(\tau)(W_k^{(\sigma)}l_k^{(\sigma)})^2\\
\times&\sin \left[\nu_k^{(\sigma)}\left(\tau^{\prime}-\tau\right)\right] \mathrm{d} \tau^{\prime} \mathrm{~d} \tau\Bigg]
\end{aligned}
\end{equation}
The geometric phases are
\begin{equation}
\begin{aligned}
\varphi^{(\sigma)}(t_g)=\vec{c}^T P^{(\sigma)}\vec{c}
\end{aligned}
\end{equation}
The residual phase for entanglement is
\begin{equation}
\begin{aligned}
\Delta\varphi=&\varphi^{(00)}(t_g)+\varphi^{(11)}(t_g)-\varphi^{(01)}(t_g)-\varphi^{(10)}(t_g)\\
=&\vec{c}^T (P^{(00)}+P^{(11)}-P^{(01)}-P^{(10)})\vec{c}
\end{aligned}
\end{equation}
We denote $P=P^{(00)}+P^{(11)}-P^{(01)}-P^{(10)}$. So, we have
\begin{equation}
\begin{aligned}
\Delta\varphi=\vec{c}^T P\vec{c}
\end{aligned}
\end{equation}
For real vector $\vec{c}$ and real matrix $P$, the relation $\vec{c}^T P\vec{c}=\vec{c}^T P^T\vec{c}$ is ensured. So, we can transform as $P\rightarrow\frac{1}{2}(P+P^T)$
to make sure $P$ is symmetric and has all real eigenvalues. Then, we diagonalize $P$ as
\begin{equation}
\begin{aligned}
P=R
\begin{bmatrix}
\lambda_1 & \\ 
&\lambda_2 &\\
& & \ddots
\end{bmatrix}
R^{-1}
\end{aligned}
\end{equation}
with $\lambda_j$ being the eigenvalues of $P$. And we have
\begin{equation}
\begin{aligned}
\Delta\varphi=\vec{c}^TR
\begin{bmatrix}
\lambda_1 & \\ 
&\lambda_2 &\\
& & \ddots
\end{bmatrix}
R^{-1}\vec{c}
\end{aligned}
\end{equation}
We can sort the eigenvalues to make the $|\lambda_1|$ to be the largest. Then, we have $|\Delta\varphi|_{max}=|\lambda_1|$ and the optimal combination coefficients are
\begin{equation}
\begin{aligned}
\vec{c}_{opt}=R\begin{bmatrix}
1\\ 
0\\
0\\
\vdots\\
\end{bmatrix}
\end{aligned}
\end{equation}
Thus, the optimal waveform can be derived as $\vec{f}_{opt}=b\cdot\vec{c}_{opt}$.
\subsection{Deriving best waveform for multi Fourier components method}
The method to optimize $|\Delta\varphi|$ for arbitrary normalized vector $\vec{A}$ in the null space of the matrix in Eq. \eqref{sinmatrixform} is the same as that in appendix A. We only need to rewrite the matrix $P^{(\sigma)}$ whose size is $2n_f\times 2n_f$
\begin{equation}
\begin{aligned}
P^{(\sigma)}=
\begin{bmatrix}
P^{(\sigma)(ss)}&P^{(\sigma)(sc)}\\
P^{(\sigma)(cs)}&P^{(\sigma)(cc)}
\end{bmatrix}
\end{aligned}
\end{equation}
The 4 part of the matrix can be expressed as
\begin{equation}
\begin{aligned}
P_{k_1, k_2}^{(\sigma)(ss)}&=\sum^k\Bigg[\int_{t_0}^t \int_{t_0}^{\tau^{\prime}}\sin(\omega_{k_1}\tau)\sin(\omega_{k_2}\tau^{\prime}) (W_k^{(\sigma)}l_k^{(\sigma)})^2\\
&\times\sin\left[\nu_k^{(\sigma)}\left(\tau^{\prime}-\tau\right)\right] \mathrm{d} \tau^{\prime} \mathrm{~d} \tau\Bigg]\\
P_{k_1, k_2}^{(\sigma)(sc)}&=\sum^k\Bigg[\int_{t_0}^t \int_{t_0}^{\tau^{\prime}}\sin(\omega_{k_1}\tau)\cos(\omega_{k_2}\tau^{\prime}) (W_k^{(\sigma)}l_k^{(\sigma)})^2\\
&\times\sin\left[\nu_k^{(\sigma)}\left(\tau^{\prime}-\tau\right)\right] \mathrm{d} \tau^{\prime} \mathrm{~d} \tau\Bigg]\\
P_{k_1, k_2}^{(\sigma)(cs)}&=\sum^k\Bigg[\int_{t_0}^t \int_{t_0}^{\tau^{\prime}}\cos(\omega_{k_1}\tau)\sin(\omega_{k_2}\tau^{\prime}) (W_k^{(\sigma)}l_k^{(\sigma)})^2\\
&\times\sin\left[\nu_k^{(\sigma)}\left(\tau^{\prime}-\tau\right)\right] \mathrm{d} \tau^{\prime} \mathrm{~d} \tau\Bigg]\\
P_{k_1, k_2}^{(\sigma)(cc)}&=\sum^k\Bigg[\int_{t_0}^t \int_{t_0}^{\tau^{\prime}}\cos(\omega_{k_1}\tau)\cos(\omega_{k_2}\tau^{\prime}) (W_k^{(\sigma)}l_k^{(\sigma)})^2\\
&\times\sin\left[\nu_k^{(\sigma)}\left(\tau^{\prime}-\tau\right)\right] \mathrm{d} \tau^{\prime} \mathrm{~d} \tau\Bigg]
\end{aligned}
\end{equation}
Using the same linear algebra methods as in appendix A, we can derive the optimal amplitude of the sine and cosine waves $\vec{A}_{opt}$. Thus, the best waveform is acquired.
\clearpage
\nocite{*}
\bibliography{shuttle_gate}
\end{document}